\documentclass[a4paper]{jpconf}
\usepackage{graphicx}
\begin{document}

\newcommand{\lag}{\mathcal{L}}
\newcommand{\TeV}{\,\mathrm{TeV}}
\newcommand{\GeV}{\,\mathrm{GeV}}
\newcommand{\MeV}{\,\mathrm{MeV}}
\newcommand{\keV}{\,\mathrm{keV}}
\newcommand{\eV}{\,\mathrm{eV}}

\title{Self-tuning of the cosmological constant\footnote{Based on talks presented at BEYOND 2010, Cape Town, South Africa, 1-6 Feb. 2010, The 6th PATRAS Workshop on Axions, WIMPs,  and WISPs, Univ. Zurich, 5-9 July 2010 and the 16th PASCOS, Valencia, Spain, 19-23 July 2010.}}

\author{Jihn E. Kim}

\address{Department of Physics and Astronomy and Center for
Theoretical Physics,\\
Seoul National University, Seoul 151-747, Korea}

\ead{jekim@ctp.snu.ac.kr}

\begin{abstract}
Here, I discuss the cosmological constant problems, in particular paying attention to the vanishing cosmological constant.
There are three cosmological constant problems in particle physics. Hawking's idea of calculating the probability amplitude for our Universe is peaked at $\overline{\Lambda}=0$ which I try to obtain after the initial inflationary period using a self-tuning model. I review what has been discussed on the Hawking type calculation with $H^2$ Lagrangian, and present a (probably) correct way to calculate the amplitude, and show that the Kim-Kyae-Lee self-tuning  model allows a finite range of parameters for the $\overline{\Lambda}=0$ to have a singularly large probability, approached from the AdS side.
\end{abstract}

%\keywords{Cosmological constant; Self-tuning; Brane scenario; Probability amplitude.}
%%%%%%%%%%%%%%%%%%%%%%%%%%%%%%%%%%%%%%%%%%%%%%%%%%%%%%%%%%%%%%%%%%%%%%%%%%%

%%%%%%%%%%%%%%%%%%%%%%%%%%%%%%%%%%%%%%%%%%%%%%%%%%%%%%%%%%%%%%%%%%%%%%%%%%%%%%
\section{Introduction}\label{sec1:Intro}
\vskip 0.3cm
Recently, Nicoli commented that the Einstein equation is inconsistent from the outset \cite{Nicoli10} because the left-hand side (LHS) and the right-hand side (RHS) of the equation
\begin{equation}
R_{\mu\nu}-\frac12 Rg_{\mu\nu}= 8\pi G_{N} T_{\mu\nu}\label{eq:Einstein}
\end{equation}
are logically different. The LHS is exactly determined by the continuous space-time geometry. It is a classical concept on the geometry. On the other hand, the RHS is contributed by quanta of particles which have roots in the probabilistic interpretation of quantum mechanics in addition to the continuous vacuum energy if present. Any discussion in the Planck era is speculative and not yet well formulated as Kolb and Turner stress in the last Chapter of \cite{KolbTurner}. To answer this question satisfactorily, we must have a good quantum theory of gravity. So, I must admit that any solution of the cosmological constant (CC) problem may be incomplete at present, and what I present here is the CC solution idea just by replacing $4\pi G_NT_{\mu\nu}$ on the RHS of (\ref{eq:Einstein}) by a classically interpretable $\Lambda g_{\mu\nu}$. At present, we do not know whether we are in a situation similar to classical electrodynamics in the early 20th century where the stability of an electron in Hydrogen needed a continuous distribution of positive charges, which was the Thompson model of Hydrogen. With quantum mechanics, we became to understand Hydrogen in terms of the Rutherford model. So, the CC problem may have a root in a completely different domain of physics on which we may be unfamiliar now. In this sense, we may look for any reasonable solution of the problem and it is welcome if it is not contradicted outrageously from the present perspective on the fundamental physics. In this spirit I present an idea with a specific action toward a CC solution. But there exists one nagging question on the proposal: why such an action? It may be like the stability problem of the Rutherford model with the 19th century eye of classical electrodynamics.

\begin{figure}[h]
\hspace{7pc}\includegraphics[width=10cm]{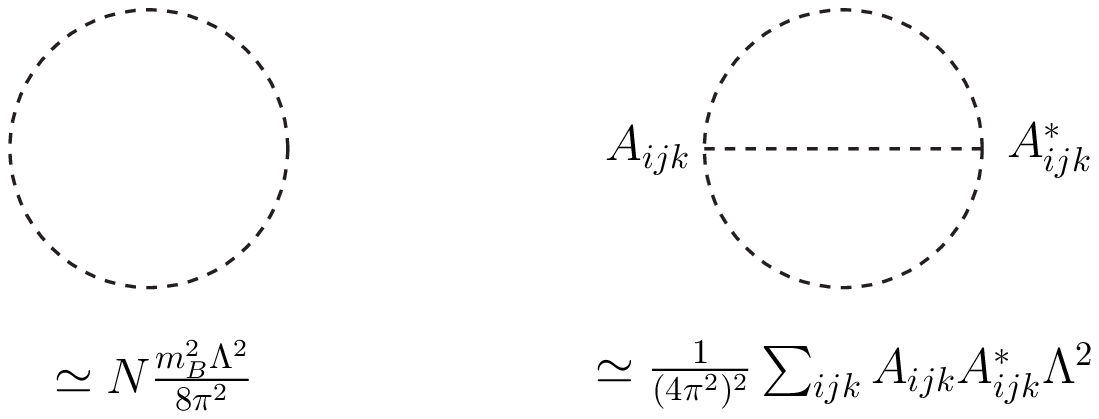}
\caption{\label{fig:loops} Loop corrections to the CC.}
\end{figure}

The CC was introduced almost ninety-three
years ago by Einstein \cite{EinsteinCC}. Then, much later when the spontaneous symmetry breaking is widely discussed in the standard model, Veltman commented that the vacuum energy arising in spontaneous symmetry breaking adds to the CC \cite{Veltman75}, basically raising a question on the naturalness of setting it to zero.
Even before considering the tree level CC, the loop
correction to the vacuum energy was a problem since the
early days of quantum mechanics. Any mode contributes $\frac12\hbar\omega$ to the vacuum energy. In the CC discussion here, we will not rely on the probably-already-happened anthropic arguments \cite{Anthropic}. So, we consider the CC generically, at the tree and also at loop levels unless the figures of Fig. \ref{fig:loops} are forbidden. For example, the LHS figure of Fig. \ref{fig:loops} corresponds to $\frac12\hbar\omega$ per mode and the RHS figure is the two-loop vacuum energy arising from the $A$-terms in supergravity \cite{CKN94}. There can be many more diagrams contributing to the vacuum energy at the loop levels.

\begin{figure}[h]
\begin{minipage}{14pc}
\includegraphics[width=15pc]{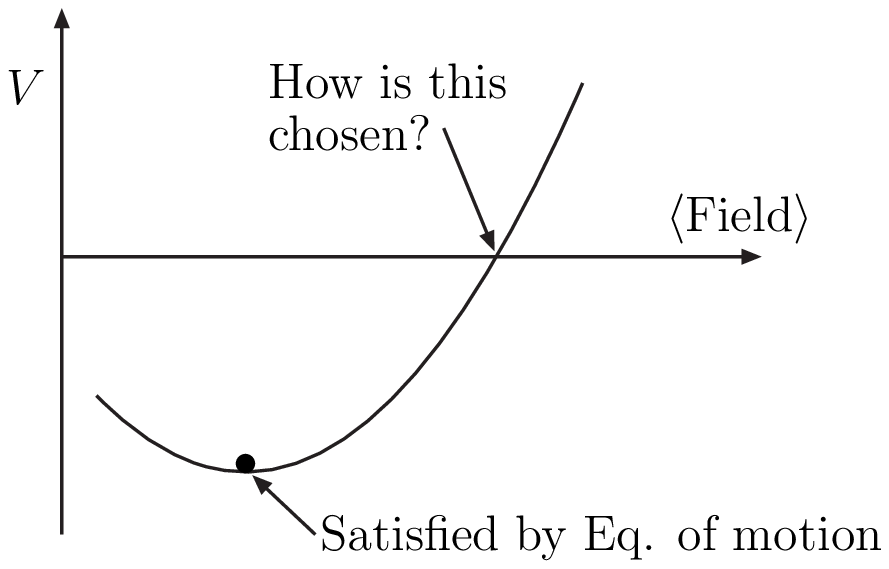}
\caption{\label{fig:potential} A form of the potential energy in terms of CC.}
\end{minipage}\hspace{6pc}%
\begin{minipage}{12pc}
\includegraphics[width=14pc]{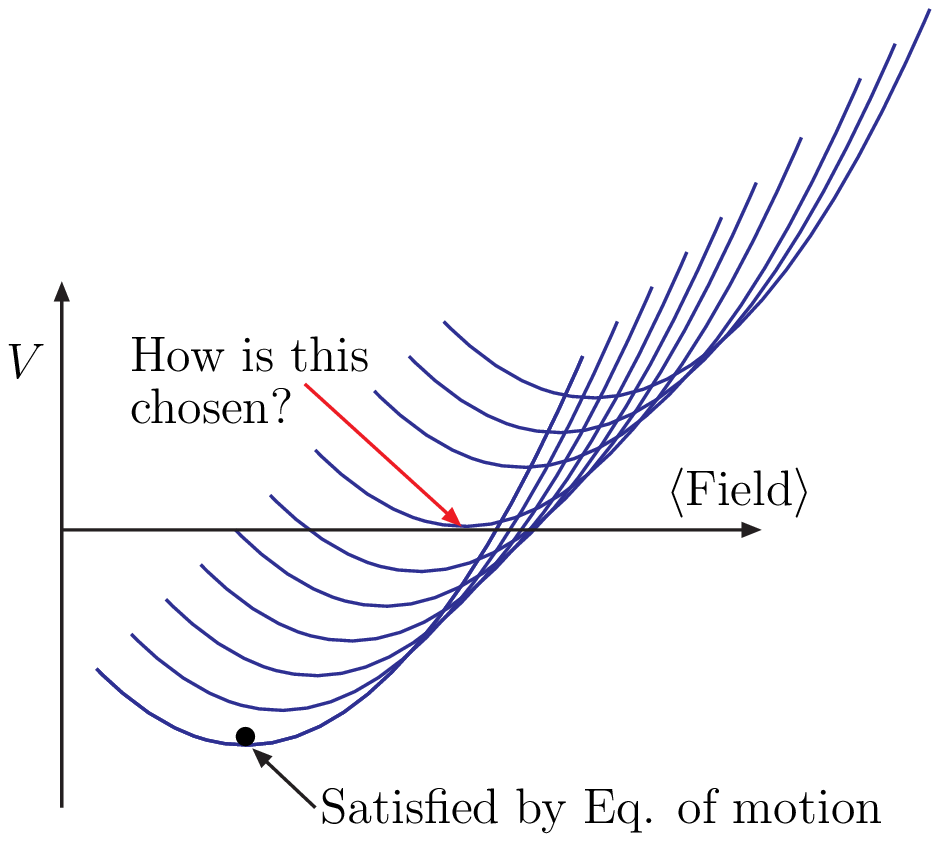}
\caption{\label{fig:varpoten} A form of the potential energy with multiple solutions.}
\end{minipage}
\end{figure}

If a symmetry is present in changing the CC $\Lambda$,
one may try a scalar potential of Fig. \ref{fig:potential} where the vertical axis corresponds to the CC. The vanishing CC is the point which the arrow points to and the vacuum is the point where the bullet is located. As Fig. \ref{fig:potential} shows, in general the vanishing CC point does not corresponds to the bullet where the equation of motion is satisfied. So, a solution is not easily realizable in four space-time dimensions (4D). In addition, the CC problem must also take into account the result of spontaneous symmetry breaking \cite{Veltman75}. The vacuum energy scale at the interesting particle physics scales are generally much greater than the current limit of $(\rm 0.002 \ eV)^4$. Thus, the CC is a serious fine-tuning problem. In 4D, we do not find any symmetry such that the CC is forbidden. Note, however, the recent tries of scale invariance and brane statistical search \cite{Shaposh08}.

Another question is at which energy scale and {\it temperature}, the CC is required to vanish. There exists a hierarchy of mass scales in particle physics:
\begin{itemize}
\item  {\rm Planck~ scale:}~   $2.44 \times 10^{18}$ GeV
\item        {\rm   GUT~ scale:} ~    $ (2-3) \times 10^{16}$ GeV
\item  {\rm Inflation~scale:}~$\simeq 10^{16}$ GeV, {\rm down~ to~ the~EW~scale}
\item   {\rm  Axion~ scale: } $~\simeq  10^{12} - 10^{11}$ GeV
\item    {\rm  Hidden~sector~ scale:}~  $10^{15}$ GeV \cite{Kim09Grav},  {\rm down~ to~ $10^6$ GeV \cite{Dine93}}
\item  {\rm  EW~ scale:}~ 100 GeV
\item    {\rm  QCD~ scale:}~      1 GeV
\item   {\rm  Nuclear~ physics~ scale:}~ 10 MeV
\item    {\rm  Electron~ mass~ scale:}~ 0.5 MeV
\item    {\rm  Neutrino~ mass~ scale:}~ $\rm 0.05\ eV$
\item    {\rm  Accelerating~ universe~ scale:}~ $\rm(0.002\ eV)^4$
\item    {\rm  Temperature~ of~ the~ CMB:}~ $\rm 10^{-4}\ eV$
\end{itemize}
For example, even though we suppose to have a CC solution at an EW scale, still $10^{-60}$ fine-tuning is required. It is a hierarchy among the parameters in the Lagrangian. Since the CC is dimension 4 parameter, it has the most severe fine-tuning problem in the Lagrangian. Thus, finding a solution of the CC problem is very difficult and is hence challenging. Therefore, any simple idea toward a solution is welcome at this stage if it is not obviously wrong.

%%%%%%%%%%%%%%%%%%%%%%%%%%%%%%%%%%%%%%%%%%%%%%%%%%%%%%%%%%%%%%%%%%%%%%%%%
\section{Probability amplitude}\label{sec2:Probamp}
\vskip 0.3cm
When we consider quantum mechanics, we talk in terms of the probability
amplitude: The initial state $|I\rangle$ transforming to a final state  $|F\rangle$. In this spirit, Baum \cite{Baum83} and Hawking \cite{Hawking84} considered the Euclidian action, only with the Ricci scalar $R$ and the CC. The method touches upon the quantum gravity, by calculating the path integral in the Euclidian spacetime. The path integral deals with all possible paths and in the $\hbar=0$ limit it is dominated by the classical equations of motion. If the topology change of the metric is considered, we must know the full gravity equation such as those including wormhole solutions \cite{GS88}, in which case an exponential of exponential function was obtained \cite{Coleman88}. For simplicity, we do not delve into an exponential of an exponential. Without the topology change which may be a reasonable assumption much below the Planck scale, this behavior of an exponential of exponential has not been known. The Euclidian action integral which we consider first from \cite{Baum83,Hawking84} has the following form
\begin{eqnarray}
e^{-\tilde I_E}=e^{3\pi M_P^2/\Lambda}.
\end{eqnarray}
So, $\Lambda=0^+$ dominates the action integral, which is interpreted as the probability for  $\Lambda=0^+$ is close to 1. But, there are questions regarding to this Baum-Hawking solution. Hawking \cite{Hawking84} states,
$\lq\lq$My proposal requires that a variable effective CC be
genarated in some manner. One possibility would be to
include the values of the CC in the variables that are
integrated over in the path integral."
For this purpose, Ref. \cite{Hawking84} explicitly considered in terms of $A_{\mu\nu\rho}$ (or the field strength $H_{\mu\nu\rho\sigma}$) which however is not a dynamical field in 4D.

In this scenario, the needed quantity to calculate is the action integral. Even if we understand the CC in this way, there exists another questions such as
\begin{itemize}
\item How do we assign the initial state?
\item How does the needed primary inflation come about in this scenario?
\item How does it fit to the current dark energy?
\end{itemize}
So the CC solution needs to answer the other two CC related questions also and furthermore needs an argument on, ``What was the proper initial condition of our Universe?"

The existing idea of Hawking in terms of $H_{\mu\nu\rho\sigma}$ with
no kinetic energy term cannot explain all the above questions. We must introduce the kinetic term with the potential shape given as that of Fig. \ref{fig:varpoten} so that the point where the equation of motion is satisfied can be the point with $\Lambda=0$.
For example, $H_{\mu\nu\rho\sigma}$ can achieve this but without the kinetic energy term in 4D. If we want to use  $H_{\mu\nu\rho\sigma}$ field, we must work beyond 4D.

%%%%%%%%%%%%%%%%%%%%%%%%%%%%%%%%%%%%%%%%%%%%%%%%%%%%%%%%%%%%%%%%%%%%%%%%%%%%%
\section{A self-tuning solution with a brane}\label{sec3:Selftuning}
\vskip 0.3cm
In 2000, self-tuning solutions have been tried in the RS type
models in 5D. Here, I just mention the initial try and its failure.
In an RS-I type model, Ref. \cite{Self1} tried to show that SM fields living at the SM brane located at $x^5=0$ do not change, via the loop corrections at the brane, the CC solution of the bulk. Here, the bulk action is fixed with specific magnitude of coupling,
\begin{eqnarray}
I_5=\int d^5x\sqrt{-g}\left[R-\frac43(\partial \phi)^2-\Lambda e^{a\phi} \right]\\
+\int d^4x\sqrt{-g_4}\left[-Ve^{b\phi} \right]_{x^5=0}.
 \label{eq:Ark}
 \end{eqnarray}
In a sense, the vacuum energy of the SM brane is cured. However, the specific form of the above bulk action is arguable for a general CC solution. Even though we allow this procedure, still it has its own problem that a singularity is present at a point $y_s$ in the bulk. The singularity can be cured by inserting the $Z_2$ symmetric branes at $\pm y_s$ \cite{SelfForste},
\begin{eqnarray}
I_5+\int d^4x\sqrt{-g_4}\left[-V_+e^{b_+\phi} \right]_{x^5=+y_s}
+\int d^4x\sqrt{-g_4}\left[-V_-e^{b_-\phi} \right]_{x^5=-y_s}.
\label{eq:Forste}
\end{eqnarray}
Then, to cancel the contribution of the SM brane of Eq. (\ref{eq:Ark}), one must fine-tune the CC contribution from the singularities of Eq. (\ref{eq:Forste}) \cite{SelfForste}.
Again a fine-tuning is needed: $\Lambda_{4D}=E_0+E_++E_-=0$, leading to a fine tuning between $V_+,b_+,V_-$, and $b_-$. Furthermore, there exists the no-go theorem under some plausible conditions such that one employs the usual kinetic energy term and assumes the Lorentz symmetry and the existence of 4D gravity for a large distance separation \cite{Csaki00}.

Here, we try to go beyond the above set-up. Namely, we do not specify the bulk action in terms of specific constants. Instead, we allow non-standard kinetic energy term to avoid the no-go theorem. In our discussion we will distinguish $\Lambda$'s, depending where it originates, the barred ones and the rest:
\begin{eqnarray}
\overline{\Lambda}={\rm obtained~ from}~ g_{\mu\nu}\nonumber\\
\Lambda={\rm obtained~ from}~ T_{\mu\nu}.\nonumber
\end{eqnarray}
In this spirit, there exists one self-tuning model by Kim, Kyae and Lee (KKL) \cite{KKL00}. The KKL model is worked out in the Randal-Sundrum II type model \cite{RSII}, with a nonstandard kinetic energy term of the antisymmetric field strength $H_{MNPQ}$: $\sim 1/H^2$ \cite{KKL00},
\begin{eqnarray}
-I_E &=\int d^5x_E \sqrt{g_{(5)}}\left(\frac12 R_{(5)}- \frac{2\cdot 4!}{H^2}-\Lambda_b -\Lambda_1\delta(y) \right)= \int dy\int d^4x_E \nonumber\\
  &\Big\{ -\Psi^4\Lambda_1\delta(y)+\frac12 R\Psi^2
+ 4\Psi^3\Psi''+6\Psi^2(\Psi')^2 +\frac{2\cdot 4!\Psi^4}{H^2}-\Psi^4\Lambda_b
 \Big\}
\end{eqnarray}
where the metric is taken as $ds^2=\beta^2(y)\eta_{\mu\nu}dx^\mu dx^\nu+dy^2$ with the signature $\eta_{\mu\nu}={\rm diag.(-1,+1,+1,+1)}$.
The kinetic energy term with $H^2$ is not developing a VEV in the low energy theory, i.e. in the long wavelength limit $(\partial_\mu A^{\nu\rho\sigma})\to 0$. Fortunately, however, we allow the bare CC in the bulk. So, even with $\langle H^2\rangle=0$, $H^2$ can be moved to the denominator,
$$
\frac{1}{H^2},\quad {\rm with}~\langle H^2\rangle\ne 0.
$$
The field equation and the Bianchi identity are satisfied with
\begin{eqnarray}
\partial_M\left(\sqrt{g} \frac{H^{MNPQ}}{H^4}\right)=0,\quad
\epsilon^{RMNPQ}[\sqrt{g_{(5)}}H_{MNPQ}]=0~.
\end{eqnarray}
With this Lagrangian, there exists a self-tuning solution
\begin{eqnarray}
\beta(y)=\left(\frac{k}{a}\right)^{1/4}\frac{1}{ (4k|y|+c_0)^{1/4}}~.
\end{eqnarray}
But there are nearby dS and AdS solutions also \cite{KKL00}. It is easy to show the existence of the nearby dS and AdS solutions by trying a small $c(y)$ in $Y\equiv \beta(y)^4 =A[{\rm sech} (ky+c_0)+c(y)]$.  The equation satisfied by $Y$ is
\begin{eqnarray}
-\frac{1}{4}Y''=3\overline{\Lambda}\sqrt{Y}+\frac{2m^2 k^2}{3}Y
-\frac{8}{3h}Y^3-\frac{\Lambda_1}{3}\delta(y)Y~
\end{eqnarray}
from which one can fix for $\Lambda_b=-m^2k^2$,
\begin{eqnarray}
m^2=\frac38,\quad \frac{16A^2}{3h}k^2 ~.
\end{eqnarray}
So, the question is, $\lq\lq$How does one choose the flat one?"

%\vskip 0.2cm
%%%%%%%%%%%%%%%%%%%%%%%%%%%%%%%%%%%%%%%%%%%%%%%%%%%%%%%%%%%%%%
\section{The initial state after inflation}\label{sec:Initial}

%%%%%%%%%%%%%%%%%%%%%%%%%%%%%%%%%%%%%%%%%%%%%%%%%%%%%%%%%%%%%%%%%%%%%
\vskip 0.3cm
\subsection{The wave function of the universe}\label{subsec:Wavefn}
\begin{figure}[h]
\begin{minipage}{12pc}
\includegraphics[width=16pc]{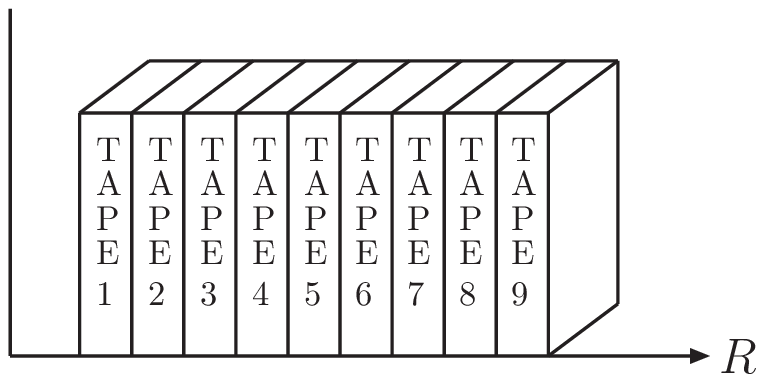}
\caption{\label{fig:Tapes} A stack of tapes in the video room. The picked tape is run with the classical Einstein equation. Different tapes contain different fundamental constants.
}
\end{minipage}\hspace{8pc}%
\begin{minipage}{14pc}
\includegraphics[width=16pc]{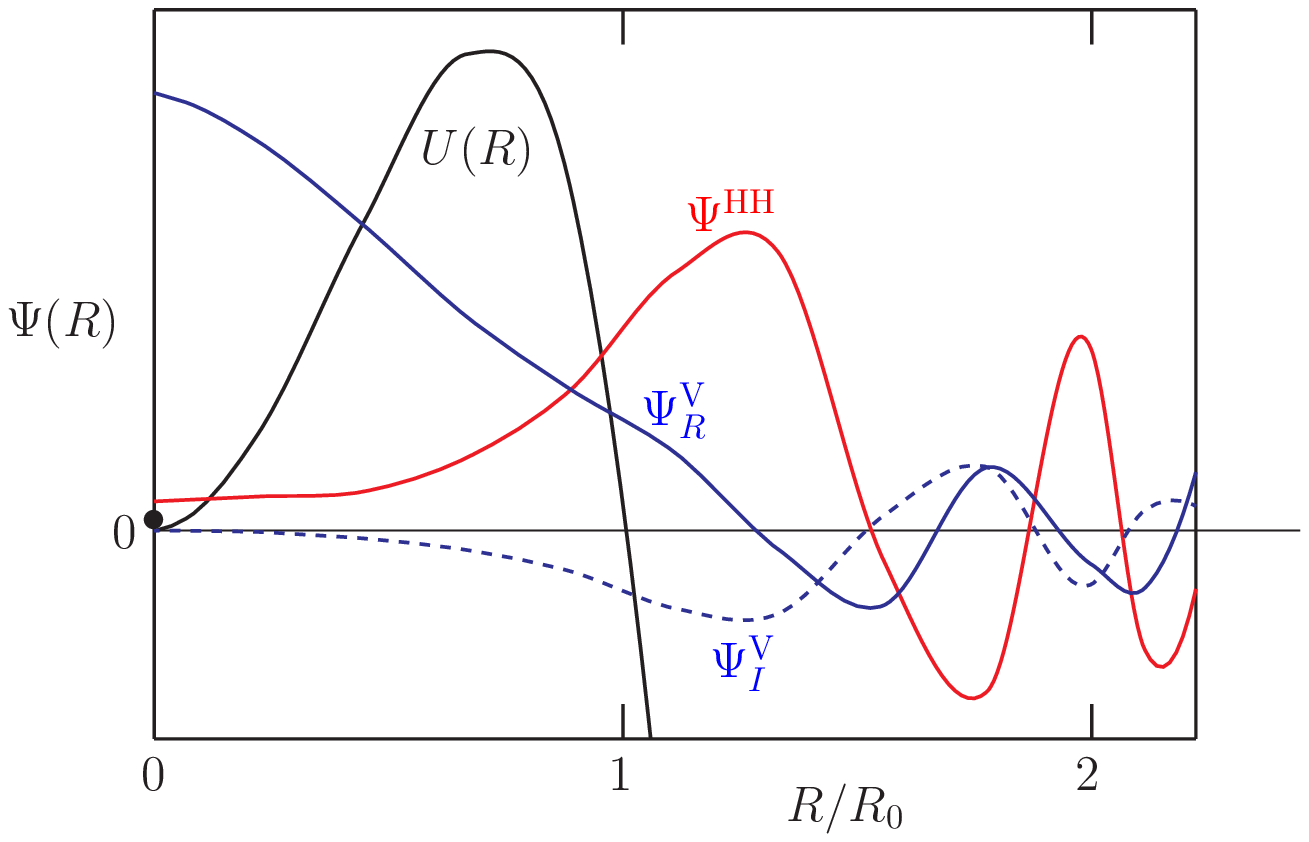}
\caption{\label{fig:Wavefunc} The universe wave functions based on the HH boundary condition and the Vilenkin boundary condition \cite{KolbTurner}. The shape of a potential $U(R)$ is also shown. The HH wave function is real while the Vilenkin wave function takes the outgoing wave only whose real and imaginary parts are shown here.}
\end{minipage}\hspace{8pc}%
\end{figure}

\hskip 0.5cm
One way of doing quantum cosmology is to solve the Wheeler-DeWit equation \cite{WDeq67} with an appropriate boundary condition. The obtained wave function of the Universe is independent of time. Following the terminology of Nicoli \cite{Nicoli10}, it is a videotape containing all the information of the universe as cartooned in Fig. \ref{fig:Tapes}. There are many videotapes satisfying the Wheeler-DeWit equation. The probability to obtain a certain videotape is given by the wave function of the universe. If one obtains a videotape, he can run it in a film motion-picture projector with a certain definition of time to see how the videotape, containing all the fundamental constants of physics, evolves. Different tapes contain different fundamental constants. Of course, the CC is contained in the videotape. In this running of the videotape, he is in the classical regime where the classical Einstein equation is enough to run the film.

After Hawking proposed calculating the probability amplitude based on the Wheeler-DeWit equation \cite{WDeq67} with the Hartle-Hawking (HH) boundary condition \cite{HHbound83}, Vilenkin suggested the outgoing wave boundary condition \cite{WiikRot}, as shown in Fig. \ref{fig:Wavefunc}. The HH wave function is real and the Vilenkin wave function is an outgoing wave whose real and impaginary parts are shown schematically in Fig. \ref{fig:Wavefunc}.  This is well summarized in Page 460 of Ref. \cite{KolbTurner}. The customary Wick rotation direction is for $E>0$ while in the Wheeler-DeWit equation the gravitational energy is negative and hence the Wick rotation has to be to the opposite direction \cite{WiikRot}, and the probability prediction becomes completely opposite to that of Hawking.

But here we follow Hawking's convention and his interpretation for the moment. Then, our result can be appropriately reinterpreted, following  \cite{WiikRot}. But our boundary condition is different from the HH or Vilenkin boundary conditions.

%%%%%%%%%%%%%%%%%%%%%%%%%%%%%%%%%%%%%%%%%%%%%%%%%%%%%%%%%%%%%%%%%%%%%
\subsection{The initial state in the brane scenario} \label{subsec:Braneinitial}
\begin{figure}[h]
\begin{minipage}{14pc}
\hspace{2pc}\includegraphics[width=10pc]{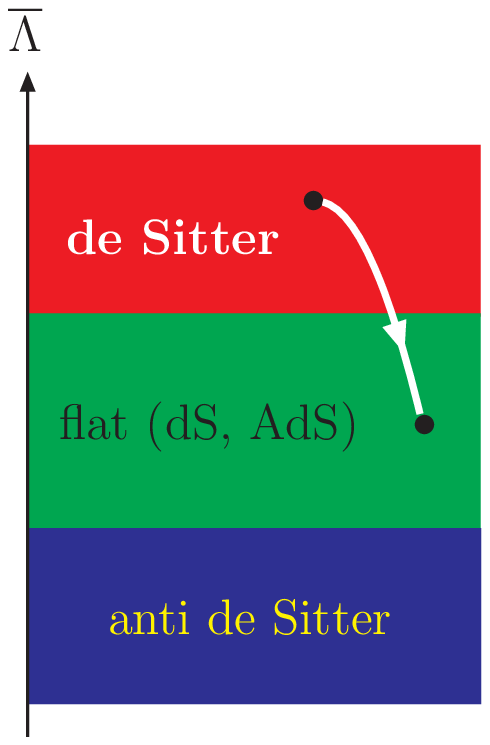}
\caption{\label{fig:InflStage} The initial inflation takes place for $|\Lambda_1|>\sqrt{-6\Lambda_b}$.}
\end{minipage}\hspace{4pc}%
\begin{minipage}{12pc}
\includegraphics[width=16pc]{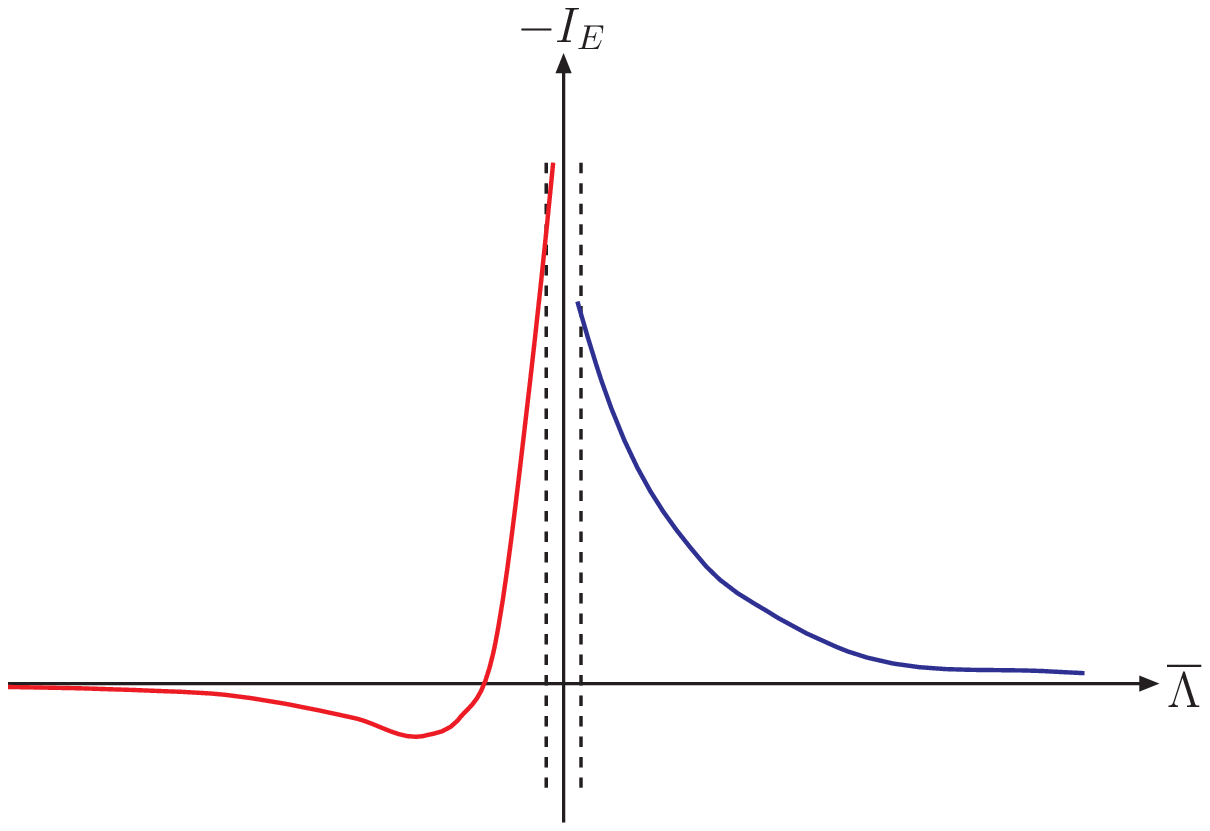}
\caption{\label{fig:CCprob} A schematic behavior of the probability amplitude in self-tuning models.}
\end{minipage}
\end{figure}

\hskip 0.5cm
This idea of quantum cosmology must be recast with extra dimensions and branes. The KKL solution has a remarkable property as noted in \cite{KimdSInfl} that the vanishing CC solution is not allowed for the parameter range of
\begin{eqnarray}
 |\Lambda_1|\ge \sqrt{-6\Lambda_b}\label{eq:LambdS}
\end{eqnarray}
since the boundary condition at the brane $(\beta'/\beta)_{y=0^+}= -\Lambda_1/6$  cannot be satisfied.  This situation is depicted in Fig. \ref{fig:InflStage}.

\begin{figure}[h]
\hspace{6pc}\includegraphics[width=25pc]{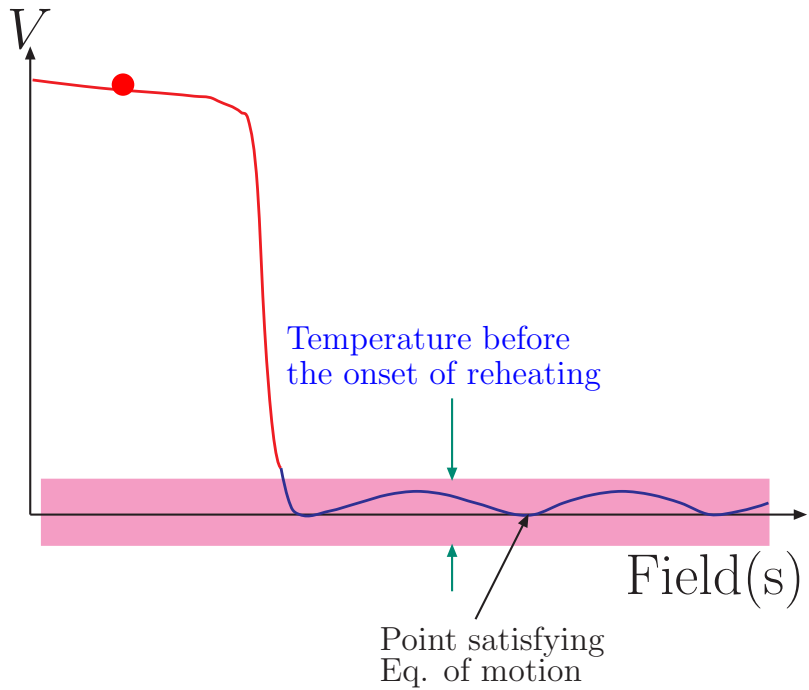}
\caption{\label{fig:InflQuint} The initial inflation and the recent acceleration. The CC solution is swamped in the lavender band and
the wave function of the universe is not exactly choosing the equilibrium point.}
\end{figure}

Therefore, we argue that for a finite range of parameter of $\Lambda_1$ satisfying (\ref{eq:LambdS}) the inflation period continues. The particle physics action at the brane may change $\Lambda_1$ such that it does not satisfy (\ref{eq:LambdS}); in the central region of Fig. \ref {fig:InflStage} all the possibilities are open, the flat, dS and AdS solutions. Then the inflationary period might end, but the important question is what is the probability to choose the flat solution. Here comes the probability calculation. In the KKL model, we show that the probability shape is given as the RHS of Fig. \ref{fig:InflQuint}.

The number of e-foldings during inflation is about 60, or the scale factor increases by a factor $10^{26}$. So, the brane Lagrangian for inflation is tuned to satisfy this condition. Then, the initial temperature $T_i$ drops to $10^{-26}T_i$ before the inflaton reheats up the universe. At this supercooled state before reheating, we define the initial state of the universe for the probability calculation. We require that this initial state is fuzzy enough to allow 0.002 eV vacuum energy, even if the probability calculation of Fig. \ref{fig:CCprob} dictates the vacuum energy should be zero.
Then, we require  $10^{-26}T_i$ be greater than $0.002$ eV, namely $T_i$ should be greater than $2\times 10^{14}$ GeV. For example, the height of the quintessential axion of Ref. \cite{quintax} can be of order $>(0.002 ~\rm eV)^4$ but the fuzziness of our CC solution is of order $(0.002 ~\rm eV)^4$, which is depicted in Fig. \ref{fig:InflQuint}. So, with any inflation model where the beginning of inflation starts below $2\times 10^{14}$ GeV, the fuzziness too small and we cannot satisfy the observed acceleration \cite{Perlmutter99} within our scheme of the CC solution.

We take our initial condition at the time immediately after the inflation. We picked up the matter energy being in a certain region after the inflation. So, our boundary condition is a matter energy eigenstate, amounting to a kind of quantum mechanical filtering process.

%%%%%%%%%%%%%%%%%%%%%%%%%%%%%%%%%%%%%%%%%%%%%%%%%%%%%%%%%%%%%%%%%%%%%%%%%%%
\section{Calculation of probability amplitude in the KKL model}
\vskip 0.3cm
Hawking calculated the probability amplitude from
\begin{eqnarray}
\langle \overline{\Lambda}|I\rangle\propto \int d[g]e^{-I_E[g]}\label{eq:ProbAmp}
\end{eqnarray}
and concluded that the volume integral is the dominant factor and concluded that the probability is the largest for $\overline{\Lambda}=0^+$ \cite{Hawking84}. In Hawking's case, it is not clear how the primordial inflation is taken into account.\footnote{Ref. \cite{WiikRot} tries to interpret the inflation due to the large probability amplitude for a large CC in case of the opposite sign of the Wick rotation, which is completely different from our view here.} In our case, the primordial inflation is achieved in the brane dynamics as explained in the previous section. Next, the initial state is given when $\Lambda_1$ rolls into the flat space region of Fig. \ref{fig:InflStage}. This initial state that $\Lambda_1$ sits in the flat space region is the quantum mechanical filtering process. So, to measure some state out of the state $|I\rangle$, we calculate the amplitude (\ref{eq:ProbAmp}).

If particle physics Lagrangian does not contribute to (\ref{eq:ProbAmp}), it is sufficient to consider the CC term and the $R$ term only as Hawking has done. Hawking's basic argument was the size of the Euclidian volume.
The dS space volume is finite, the flat space volume is infinite,
and the AdS space volume is even more infinite. If we consider only the $1/\,\overline{\Lambda}$ term, the AdS wins in magnitude but the sign is opposite from the dS; thus the flat space is chosen. So, even if the AdS is considered, the flat space wins if we restricts only to $1/\,\overline{\Lambda}$ term. However, as in the self-tuning model of KKL, particle physics Lagrangian contributes to the final CC in general, and the particle physics action has the $1/\,\overline{\Lambda}^{\,2}$ behavior which dominates the CC term contribution in the small $\overline{\Lambda}$ region. This may change Hawking's view completely.

If we consider the sizes of volumes, the AdS wins over the flat
even though both are infinite. For the flat volume, we take the
$\overline{\Lambda}=0$ limit of the dS case. For the AdS volume, we need to
regularize the infinity to compare different cases of $\overline{\Lambda}$'s.

Here, I discuss what has been discussed on the $H^2$ Lagrangian in the Hawking type calculation, present a (probably) correct way to calculate the amplitude, and show that the KKL self-tuning  model allows a finite range of parameters for which $\overline{\Lambda}=0^-$ has the singularly large probability \cite{Kim09}.

%%%%%%%%%%%%%%%%%%%%%%%%%%%%%%%%%%%%%%%%%%%%%%%%%%%%%%%%
\subsection{Hawking, Duff and Wu}
\vskip 0.3cm
\hskip 0.5cm
Hawking presented the first calculation, using $H^2$ term in the Lagrangian \cite{Hawking84}. Since then,
there has been discussions on which value of the $H^2$ term must be used in the action integral.  In this regard, a surface term $\epsilon^{\mu\nu\rho\sigma}H_{\mu\nu\rho\sigma}$ has been noted, which does not change the equation of motion. It turned out that it amounts to changing the sign of $H^2$ term in the action integral \cite{Duff89}, from which Duff concluded that the probability amplitude for $\overline{\Lambda}=0$ is least probable with $H_{\mu\nu\rho\sigma}$. Recently, Wu \cite{Wu08} obtained the opposite result from that of Duff. In our case also, we can consider a
surface term (due to $x$-independent $\Psi^8/H^2$), following the first order formalism of \cite{Aurelia80},
$$
-I_{E}\supset\int dy\int d^4x_E\left\{ \rho\frac{2\Psi^{8}}{H^2} \epsilon^{\mu\nu\rho\sigma}H_{\mu\nu\rho\sigma} -\frac{\Psi^{4}}{2} \rho^2\right\}
$$
which does not affect the equation of motion. If we follow Duff's method, it has the effect of changing the sign of $1/H^2$ term inside the action integral with the surface term neglected, from $\int d^5x_E \sqrt{g_{(5)}}(2\cdot 4!\Psi^4/H^2)$ to  $\int d^5x_E \sqrt{g_{(5)}}(-2\cdot 4!\Psi^4/H^2)$. So, both methods satisfy the equations of motion with the action integral,
\begin{eqnarray}
&{\rm Duff}\nonumber\\
&\downarrow\nonumber\\
-I_E =\int d^5x_E &\quad\quad\ \ \sqrt{g_{(5)}}\Big[\frac12 R_{(5)}\pm \frac{2\cdot 4!}{H^2}-\Lambda_b -\Lambda_1\delta(y) \Big]\\
&\uparrow\nonumber\\
&{\rm Wu}\nonumber
\end{eqnarray}

Thus, it raises an important question, $\lq\lq$Which method is correct?"

%%%%%%%%%%%%%%%%%%%%%%%%%%%%%%%%%%%%%%%%%%%%%%%%%%%%%%%%
\subsection{The $\alpha$-vacuum}
\vskip 0.3cm
\hskip 0.5cm
To our view, the confusion arises from taking a specific vacuum in their calculation \cite{Duff89,Wu08}. As in the $\theta$-vacuum of QCD, we have the $\alpha$-vacuum of antisymmetric tensor field $H_{\mu\nu\rho\sigma}$. Duff took one extremum point corresponding to $\alpha=\pi$ and Wu took another vacuum corresponding to $\alpha=0$, and they obtained different results even though both satisfied equations of motion.

As far as the $\alpha$-vacuum is concerned, the discussion is parallel whether we use $H^2$ or $1/H^2$ in the Lagrangian. So, for the notational brevity, we discuss $\alpha$-vacuum with the $H^2$ kinetic energy term.

For two antisymmetric indices from $\mu,\nu,\rho,$ and $\sigma$, there are
six ($_4C_2=6$) independent second rank antisymmetric gauge
functions, for which $A_{\mu\nu\rho}$ transforms as
\begin{eqnarray}
A_{\mu\nu\rho}\to A_{\mu\nu\rho}-\partial_\mu \Lambda_{\nu\rho}-\partial_\nu \Lambda_{\rho\mu}-\partial_\rho \Lambda_{\mu\nu}~.
\end{eqnarray}
The gauge symmetry of the instanton solution is given by any six directions of $\Lambda_{\mu\nu}$, three for the instanton ($\Lambda_{ij}$) and three for the anti-instanton ($\Lambda_{0i}$) and the instanton action is $\int d^5x \partial_y\epsilon^{\mu\nu\rho\sigma} H_{\mu\nu\rho\sigma}$. Namely, there exist maps of
\begin{eqnarray}
S_3\to S_3.
\end{eqnarray}

There exists only one type of solution, i.e. only one Pontryagin index. For $\int d^4x H_{\mu\nu\rho\sigma}$ to be finite, $H_{\mu\nu\rho\sigma}$ should tend to $r^{-4}$ for a large $r$. In the bulk, it arises from a 2D curl in 5D (time and the internal space $y$),
\begin{eqnarray}
&\int dydx^0 ~ \vec{\nabla}\times \vec{A}=\int d\vec{s}\cdot \vec{A}=\int dydx^0 [\partial_y {\rm A}_0 -\partial_0 {\rm A}_y ]
\nonumber \\
&=\int dydx^0 \partial_y {\rm A}_0 =\int dx^0  {\rm A}_0 =\int d^4x H_{\mu\nu\rho\sigma}
\end{eqnarray}
where
\begin{eqnarray}
{\rm A}_0 &= \int d^3{\bf x}~\partial_0 A_{ijk}= \int d^3{\bf x}~H_{\mu\nu\rho\sigma}\nonumber \\
{\rm A}_y &= \int d^3{\bf x}~\partial_y A_{ijk}= \int d^3{\bf x}~H_{yijk}~.
\end{eqnarray}
So a gauge invariant instanton of size $\rho$ located at $x_0$ takes the form
\begin{eqnarray}
A_{\alpha\beta\gamma}\propto \frac{\epsilon_{\alpha\beta\gamma\mu}(x-x_0)^\mu}{(r^2+\rho^2)^2}, \quad r=|x-x_0|
\end{eqnarray}
so that $A_{\mu\nu\rho}$ is proportional to $r^{-3}$, and $H_{\mu\nu\rho\sigma}$is proportional to $r^{-4}$. The 4D integral of $H_{\mu\nu\rho\sigma}$ is represented by a kind of Pontryagin integer $n=\pm 1$. Note, on the other hand, that the instanton field of nonabelian gauge groups is of pure gauge form, so that the instanton action is $\int d^4x {\rm Tr}F\tilde F$. In nonabelian gauge theories, there are many possible gauge configurations such that the irreducible instanton solution give many possible integers for the Pontryagin index. On the other hand, in our case at hand $H_{\mu\nu\rho\sigma}$ instanton gives only $\pm 1$ for the Pontryagin index.

Now, we construct a gauge invariant $\alpha$-vacuum, following the $\theta$-vacuum construction of QCD,
\begin{eqnarray}
|\alpha\rangle = \sum_{n=-\infty}^{+\infty}|n\rangle e^{in\alpha}
\end{eqnarray}
In the $\alpha$-vacuum, after integrating out the $H^2$ field, what Duff chose is $\alpha=\pi$ and what Wu chose is $\alpha=0$. However, in the $\alpha$ vacuum, any value of  $\alpha$ is allowed, i.e. not restricted to  $\alpha=0$ and $\pi$.  As in the $\theta$-vacuum of QCD, any value of $\alpha$ is allowed in our case, and we go beyond what Duff and Wu considered.

As commented above, this $\alpha$-vacuum can be defined also with the $1/H^2$ term. We calculate the action integral for $\alpha=0$ and $\pi$ with the $1/H^2$ term and for any $\alpha$ the action integral is between them. If one makes $\alpha$ a dynamical field as the QCD axion \cite{KimRMP10}, then $\alpha$ is cosmologically settled to 0.

%%%%%%%%%%%%%%%%%%%%%%%%%%%%%%%%%%%%%%%%%%%%%%%%%%%%%%%%
\subsection{Parameters for the $\overline{\Lambda}=0$ dominance}
\vskip 0.3cm
\hskip 0.5cm
Now, we calculate the probability amplitude in the KKL model. For this, we use two Einstein equations of the bulk,
\begin{eqnarray}
&(\mu\nu):~
-3\frac{\overline{\Lambda}}{\beta^2}+3\left(\frac{\beta'}{\beta}\right)^2
+3\frac{\beta''}{\beta}=-\Lambda_b-2\cdot 4!\left(\frac{3}{H^2}\right)\\
&(55):~
-\frac{6\overline{\Lambda}}{\beta^2}+6\left(\frac{\beta'}{\beta}\right)^2
=-\Lambda_b-2\cdot 4!\left(\frac{1}{H^2}\right)
\end{eqnarray}
We integrate out the 4D space $x$ and the 5th space $y$. In this calculation the brane tension $\Lambda_1$ is also considered. For the coefficient of  $1/\,\overline{\Lambda}^{\,2}$ to be positive, the following condition on the parameters is required,
\begin{eqnarray}
\tanh(c_0){\rm sech}^2(c_0)\leq \frac{k}{3} F(c_0/k,d_m)\label{eq:positive}
\end{eqnarray}
where $F(c_0/k,d_m)$ is the result of the integration. Here, $d_m$ is the length scale defined from the parameters of \cite{Kim09}. If this condition is satisfied, the action integral $-I_E$ has the behavior shown in Fig. \ref{fig:CCprob}, and the vanishing CC is approached from the AdS side.

In the gauge invariant $\alpha$-vacuum, for the $c_0$ independent part  with the $1/H^2$ term we obtain $(3/8k)(\pi/2)$  and $(9/2k)(\pi/2)$ for $\alpha=0$ and $\pi$, respectively. Therefore, it seems that for the parameters satisfying Eq. (\ref{eq:positive}) we obtain the $\overline{\Lambda}=0$ dominance in the probability amplitude.

%%%%%%%%%%%%%%%%%%%%%%%%%%%%%%%%%%%%%%%%%%%%%%%%%%%%%%%%%%%%%%%%%%%%%%%%%%%%
\subsection{The AdS volume}
\vskip 0.3cm
\hskip 0.5cm
Finally, we comment on our method of comparing the infinite volumes of the AdS spaces.
The $n$-dimensional Euclidian space metric is given by
\begin{eqnarray}
ds^2=a^2\left(\frac{dr^2}{1-kr^2}+r^2 d\Omega^2_{n-1}\right)
=a^2f^2(\eta)\left(d\eta^2+\eta^2 d\Omega^2_{n-1}\right)\label{eq:metric}
\end{eqnarray}
where $k=0,\pm 1$ and in the second equation the Weyl transformation is used since it is simple because of the vanishing Weyl tensor in this space,
\begin{eqnarray}
\frac{dr}{\sqrt{1-kr^2}}=f(\eta)d\eta,\quad f(\eta)=\frac{r}{\eta}~ .\label{eq:WeylTr}
\end{eqnarray}
Now, let us specify to the AdS space of $k=-1$. Then, Eq. (\ref{eq:WeylTr}) is integrated to give $\ln\eta=-\sinh^{-1} (1/r)$, or
\begin{eqnarray}
f(\eta)=\frac{r}{\eta}=\frac{2}{1-\eta^2}.
\end{eqnarray}
The Ricci scalar for the metric $g'_{\mu\nu}=a^2f^2(\eta)g_{\mu\nu}$ is given by
\begin{eqnarray}
R'=a^{-2}f^{-2}\left(R-2(n-1)\nabla^2(\ln f)-(n-1)(n-2)\left(\frac{f'}{f}\right)^2\right)~ .\label{eq:Ricci}
\end{eqnarray}
Noting that $R'_{\mu\nu}=\overline{\Lambda}g_{\mu\nu}'$ or $R'=n\overline{\Lambda}$, we have $\overline{\Lambda}= -(n-1)/a^2$ in the
$n$-dimensional Euclidian AdS.
Using Eq. (\ref{eq:WeylTr}), the $n$-dimensional Euclidian AdS volume with the metric (\ref{eq:metric}) is regularized to
\begin{eqnarray}
&V_{{\rm AdS}^n} =a^n\int d^nx\left(\frac{2}{1-\eta^2}\right)^n=
a^n\int_0^1 d\eta ~\eta^{n-1}V_{S^{n-1}}\left(\frac{2}{1-\eta^2}\right)^n=(2a)^n V_{S^{n-1}} \\
&\cdot\,\frac12\int_0^1 d\xi ~\xi^{n/2-1}(1-\xi)^{-n}
=\frac12(2a)^n V_{S^{n-1}} B(n/2, 1-n)\\
&=\frac12 (2a)^n\frac{2\pi^{n/2}}{\Gamma(n/2)} \frac{\Gamma(n/2)\Gamma(1-n)}{\Gamma(1-n/2)}=\left( \frac{4(n-1)\pi}{|\overline{\Lambda}|}\right)^{n/2}
\frac{\Gamma(1-n)}{\Gamma(1-n/2)}~ .\label{eq:AdSVol}
\end{eqnarray}
Eq. (\ref{eq:AdSVol}) factored out the diverging Gamma functions, and we can compare the $\overline{\Lambda}$ dependences. For $n=4$, it diverges as $1/|\overline{\Lambda}|^2$ as $\overline{\Lambda}$ tends to zero.

%%%%%%%%%%%%%%%%%%%%%%%%%%%%%%%%%%%%%%%%%%%%%%%%%%%%%%%%%%%%%%%%%%
\section{Conclusion}
\vskip 0.3cm
In conclusion, we observed that ({\it i}) the CC problem may be understandable  in higher dimensions $D>4$, ({\it ii}) three CC problems should be addressed, and ({\it iii}) the initial state of the Universe should be defined properly after the initial inflation. A brane helps in solving the vanishing CC problem, since the loop effects of brane is not important to bulk physics. However, this idea is applicable only when there exists a self-tuning solution such as the one given in \cite{KKL00}.

We noted that the action integral for a probability calculation is dominated from the particle physics Lagrangian, and has the amplitude proportional to $\exp[\#/\,\overline{\Lambda}^{\, 2} ]$. Near $\overline{\Lambda}=0$, the AdS space can be made to be preferred. But slightly outside  $\overline{\Lambda}=0$, the dS space is preferred. With the parameters satisfying Eq. (\ref{eq:positive}),  $\overline{\Lambda}=0$ is preferred with the Hawking's direction of the Wick rotation, which may be thought not the correct Wick rotation as commented in \cite{WiikRot}. However, our maximum probability occurs from the negative energy side, which means that the gravitational energy is positive and our choice can be the correct direction of the Wick rotation. So, we must choose parameters such that  Eq. (\ref{eq:positive}) is satisfied.\footnote{Note that if we made an error in this argument, we have a freedom to choose the opposite direction for the inequality in  Eq. (\ref{eq:positive}) to meet our need.}

Also, we noted that the current acceleration should be addressed, which has not been discussed here. For this, the quintessential-axion idea may be useful \cite{quintax}, and we need the onset of inflation above $2\times 10^{14}$ GeV which is quite reasonable and hints a GUT inflation.

Our specific example presented here for the probability amplitude calculation uses the three index gauge field $A_{MNP}$ in
the KKL model \cite{KKL00}. Here, one can consider an $\alpha$-vacuum.
Then, $\alpha$ becomes a parameter in the model and any value of $\alpha$ between 0 and $\pi$ are permitted. It is like the $\theta$ parameter of QCD. We have shown that for any value of $\alpha$, there exists a finite range of parameters such that $\overline{\Lambda}=0^-$ is chosen.

If $\alpha$ is made dynamical as the QCD axion, then the probability amplitude choosing  $\overline{\Lambda}=0^-$ is at $\alpha=0$. Depending on the scale of the $H_{\mu\nu\rho\sigma}$ instanton size, there may be some cosmological interests.
\vskip 0.3cm
\noindent {\bf Acknowledgments}: I have benefitted from collaborations and discussions with B. Kyae, H. M. Lee and J.-H. Huh. I thank the organizers of PASCOS 2010, especially to C. Mu\~noz and J. Valle for the hospitality during the conference. This work is supported in part by the Korea Research Foundation, Grant No. KRF-2005-084-C00001.

\end{document}